\documentstyle[prl,aps]{revtex}
\input epsf.tex

\newcommand{\nc}{\newcommand}
\nc{\beq}{\begin{equation}}
\nc{\eeq}{\end{equation}}
\nc{\beqa}{\begin{eqnarray}}
\nc{\eeqa}{\end{eqnarray}}
\nc{\lra}{\leftrightarrow}
\def\sfrac#1#2{{\textstyle\frac#1#2}}
\nc{\sss}{\scriptscriptstyle}
{\nc{\lsim}{\mbox{\raisebox{-.6ex}{~$\stackrel{<}{\sim}$~}}}
{\nc{\gsim}{\mbox{\raisebox{-.6ex}{~$\stackrel{>}{\sim}$~}}}
\def\VEV#1{{\langle #1 \rangle}}

\begin{document}
\twocolumn[\hsize\textwidth\columnwidth\hsize\csname@twocolumnfalse%
\endcsname

\draft
\title{Supersymmetric Electroweak Phase Transition:\\
Baryogenesis versus Experimental Constraints}

\author{James M.~Cline, Guy D.~Moore}

\address{Physics Department, McGill University,
3600 University Street, Montr\'eal, Qu\'ebec, Canada H3A 2T8.}

\maketitle

\begin{abstract} We use the two loop effective potential to study the
first order electroweak phase transition in the minimial supersymmetric
standard model.  A global search of the parameter space is made to
identify parameters compatible with electroweak baryogenesis.  We
improve on previous such studies by fully incorporating squark and
Higgs boson mixing, by using the latest experimental constraints, and
by computing the latent heat and the sphaleron
rate.  We find the constraints $\tan\beta > 2.1$, $m_{\tilde t} < 172$
GeV, and $m_{h} < 107$ GeV (becoming more or less restrictive if the
heavy stop has mass less than or greater than 1 TeV).  We also find
that the change in $\tan\beta$ in the bubble wall is rarely greater
than $10^{-3}$, which severely constrains the mechanism of
baryogenesis.  
\end{abstract}

]

%
%

The most predictive theory of the origin of baryonic matter at present
is electroweak baryogenesis, since it relies on assumptions about new
physics that is being searched for at LEP and Fermilab.  For sufficient
baryogenesis, the electroweak phase transition (EWPT) must be strong;
the requirement is roughly $v/T_c > 1$, with $v$ the Higgs vacuum
expectation value and $T_c$ the critical temperature.  This is
unachievable in the standard model.  One of the main contenders for the
new physics needed is supersymmetry, whose simplest manifestation is
the Minimal Supersymmetric Standard Model (MSSM).  Much progress has
been made in the last few years toward identifying which regions of the
MSSM parameter space are compatible with electroweak baryogenesis
\cite{eqz}--\cite{lr1}.  It is agreed that a light Higgs boson and a
light top squark are needed, but estimates vary as to just how light.
One goal of the present work is to refine these estimates within the
framework of the effective potential approach.

Computing the properties of the EWPT is subtle because the
power-counting arguments for perturbation theory at zero temperature
are modified at high temperature, such that the loop expansion for the
free energy (effective potential) must be resummed.  Even after
resumming, the effective potential (EP) is not guaranteed to be
reliable at small values of $v$, due to infrared divergences coming
from the small masses of the transverse gauge bosons and the light
Higgs boson.  There are three principal ways of dealing with this
problem: (1) compute phase transition properties fully numerically, on
the lattice \cite{lr1}; (2) integrate out the heavy, nonIR-divergent
modes analytically (``dimensional reduction'') and save the lattice for
the effective theory of the dangerous light modes
\cite{ck1,laine,losada}; and (3) improve the effective potential by
computing to higher order in the loop expansion \cite{jre,bjls,dce}.

The first approach, because of its high numerical cost, is impractical
from the standpoint of exploring large regions of parameter space.
Dimensional reduction is better in this respect because it maps the
full MSSM parameter space onto a much smaller set of parameters in the
effective high-$T$ theory for the light modes, thereby reducing the
number of lattice simulations needed.  But DR has its limitations,
especially when it comes to the effects of superrenormalizable
interactions which can induce a large number of unsuppressed
high-dimension operators in the effective theory to be studied on the
lattice.  The $A_t$ and $\mu$ parameters of the MSSM come with
precisely these kinds of interactions.

On the face of it, computing the effective potential to higher order in
perturbation theory would not seem promising since the convergence of
the perturbation series is supposed to be poor.  However experience
with the standard model shows that the two-loop EP works quite well.
In the MSSM there are also indications that the two-loop EP gives
converging results, since they are in fairly good agreement with recent
lattice computations:  the lattice gives values of $v/T$ (a measure of
the strength of the transition) which are about 10\% higher than those
of the EP.  This, together with the ability it affords for quickly
combing the full MSSM parameter space, makes it worthwhile to
investigate the two-loop EP.   Furthermore, detailed properties of the
phase transition like the nucleation temperature $T_{\rm nuc}$,
sphaleron energy inside the bubbles, and the variation of the two Higgs
field VEV's inside the bubble walls, are much more readily available
from the EP than from lattice studies.

The important contributions to the two-loop EP in the MSSM have been
calculated by references \cite{jre,bjls,dce}.  The first reference
included squark mixing, while assuming the heavy Higgs bosons are
decoupled.  The third allowed for lighter Higgs bosons, but ignored
squark mixing.  We have generalized these results to incorporate both
effects.  This is desirable because most baryogenesis mechanisms rely
upon $\tan\beta$, the ratio ($v_2/v_1$) of the two Higgs VEV's,
changing from the interior to the exterior of the bubbles, which is
only possible if the heavy Higgs bosons are not decoupled.  Squark
mixing is also expected because the phase (and magnitude) of the $\mu$
parameter must not be too small, since this is the principal source of
CP violation \cite{hn}--\cite{cjk}.  The $\mu$ parameter appears in the
off-diagonal term of the stop mass matrix, while the diagonal term for
the right-handed stop must be small to get a strong enough phase
transition. Thus significant squark mixing is a possibility, which in
fact is realized (figure 2 below).

A recent development in the literature is to consider the formation of
squark condensates at the beginning of a two-stage transition, where
color and electric charge are temporarily broken  \cite{bjls,cqw2}.  To
recover our known world, the color broken state must later copiously
nucleate bubbles of the conventional SU(2)-breaking ground state of the
standard model.  If this could happen, it would considerably strengthen
the EWPT.  However, this nucleation process is heavily suppressed
\cite{kls}, leading to the same problem that killed ``old''
inflation--the CCB vacua would inflate and vastly dominate the spatial
volume of the universe.  Hence, at all temperatures above the
nucleation temperature for the electroweak broken phase, the effective
potential for the right stop must have a stable or sufficiently
metastable symmetric minimum to prevent stop condensation.  Analyzing
the one loop effective potential for the right stop, we find that
nucleation of the color broken vacuum happens whenever $M^2_{\tilde
t_R}(T) < 0.035\, g_s^4\, T^2$.  We therefore discard parameters where
this happens at any $T > T_{\rm nuc}$.

The most relevant laboratory constraints concern $m_h$, $m_{\tilde t}$
and $m_{\tilde b}$ (the lightest Higgs boson, light stop and sbottom
masses, respectively), $\Delta\rho$ (the contributions of the stops and
sbottoms to the $\rho$ parameter), and the exclusion of charge- and
color-breaking minima.  The latest experimental limit on $m_h$ depends
on $\sin^2(\alpha-\beta)$, where $\alpha$ describes the composition of
the light Higgs field $h$ through $h = \sin\alpha H_1^0 + \cos\alpha
H_2^0$ \cite{HHG}.  The 95\% c.l.\ excluded region is roughly given by
the intersection of $m_h > (69 + 19 \sin^2(\alpha-\beta))$ GeV$/c^2$
and $m_h > (76 - 11.5 \sin^2(\alpha-\beta))$ GeV$/c^2$  \cite{aleph1}.
This allowed region and our accepted Monte Carlo points are shown in
figure 1.  The limit $\sin^2(\alpha-\beta)\to 1$ corresponds to a heavy
$A^0$ boson and a standard-model-like Higgs sector, with only one light
Higgs boson.  The phase transition is typically strongest in this
regime.  

For the squark masses we use the preliminary ALEPH limit of $m_{\tilde
t} > 82$ GeV, which is left-right mixing independent, and $m_{\tilde b}
> 79$ GeV \cite{aleph2}.  Concerning the deviation in the $\rho$
parameter; the standard model value of $\Delta\rho$ (also known as
$\epsilon_1$) is already $1.5\sigma$ larger than the experimental value
\cite{abc} for $m_h\sim 100$ GeV.  We constrain the squark contribution
to $\Delta\rho$ to be less than approximately one additional standard
deviation, namely $\Delta\rho<1.5\times 10^{-3}$.  Chargino/neutralino
searches also constrain $|\mu|\gsim 100$ GeV \cite{opal}.

We search this allowed parameter space for those values that give a
strong enough phase transition.  Our criterion is that the integrated
rate of sphaleron transitions since the phase transition reduces the
baryon asymmetry by just one $e$-folding.  Writing the sphaleron rate
per unit volume as $\Gamma_s$ and the sphaleron energy as $E_{\rm
sph}$, the bound is $\Gamma_s < \Gamma_{\rm crit}$ \cite{gm}, where
\beqa
	\label{sphrate}
	\Gamma_s &\simeq& \left(\frac{v}{T_r}\right)^7 T_r^4\, 
		e^{6.9 - E_{\rm sph}(T_r)/T_r};\nonumber\\
	\Gamma_{\rm crit} &=&  \frac{14}{123} T_r^3 
	\frac{d}{dt}\ln\Gamma_s \, ,
\eeqa
all evaluated at the reheat temperature $T_r \lsim T_c$.  We find the
sphaleron energy by solving for the sphaleron configuration using the
two loop effective potential, and we multiply both $v/T_r$ and and
$E_{\rm sph}/T_r$ by a correction factor of 1.1 suggested by the
lattice results \cite{lr1}.  The frequencies of the MSSM input
parameters which pass all these cuts, as well as histograms for some
derived quantities, are shown in figure 2.

\centerline{\epsfxsize=3.5in\epsfbox{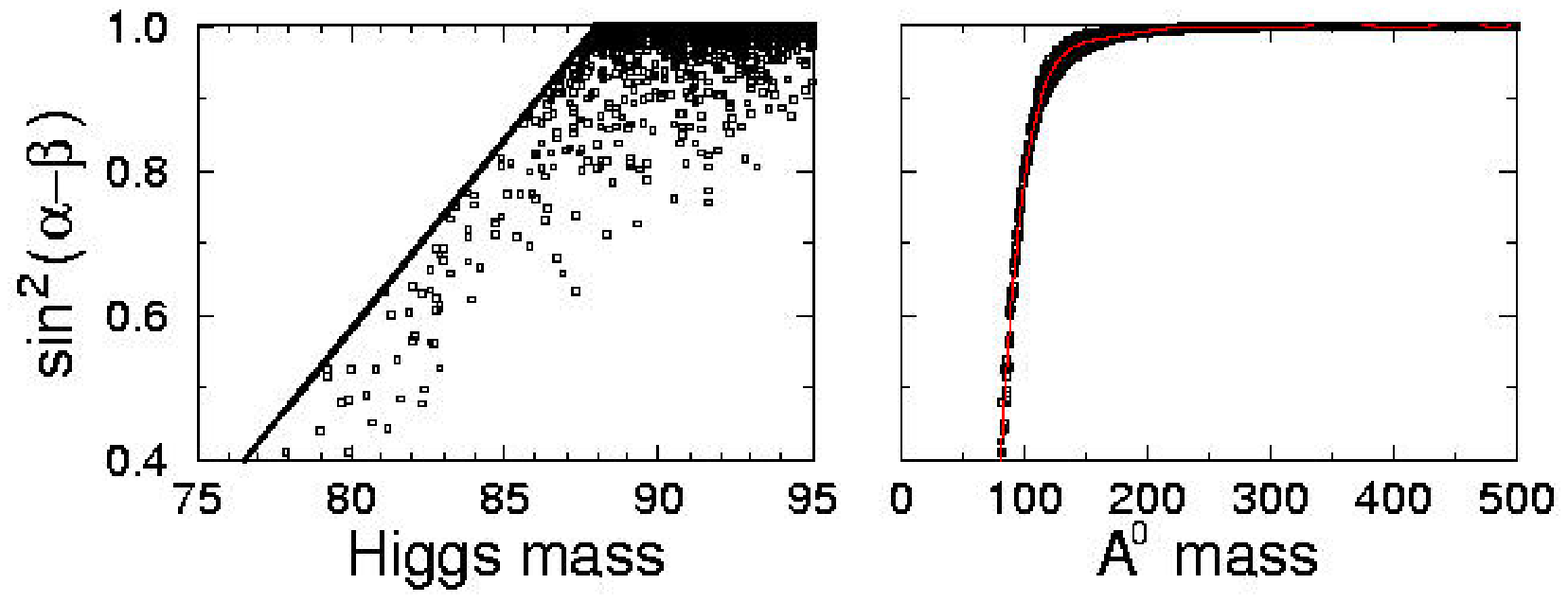}}
\vspace{-0.05in}
\noindent {\small Figure 1: (a) Experimentally allowed values of
$\sin^2(\alpha-\beta)$ and $m_h$ are to the right of the line; Monte
Carlo generated points are those consistent with electroweak
baryogenesis. (b) Relation between $\sin^2(\alpha-\beta)$ and $m_A$.}

\centerline{\epsfxsize=3.5in\epsfbox{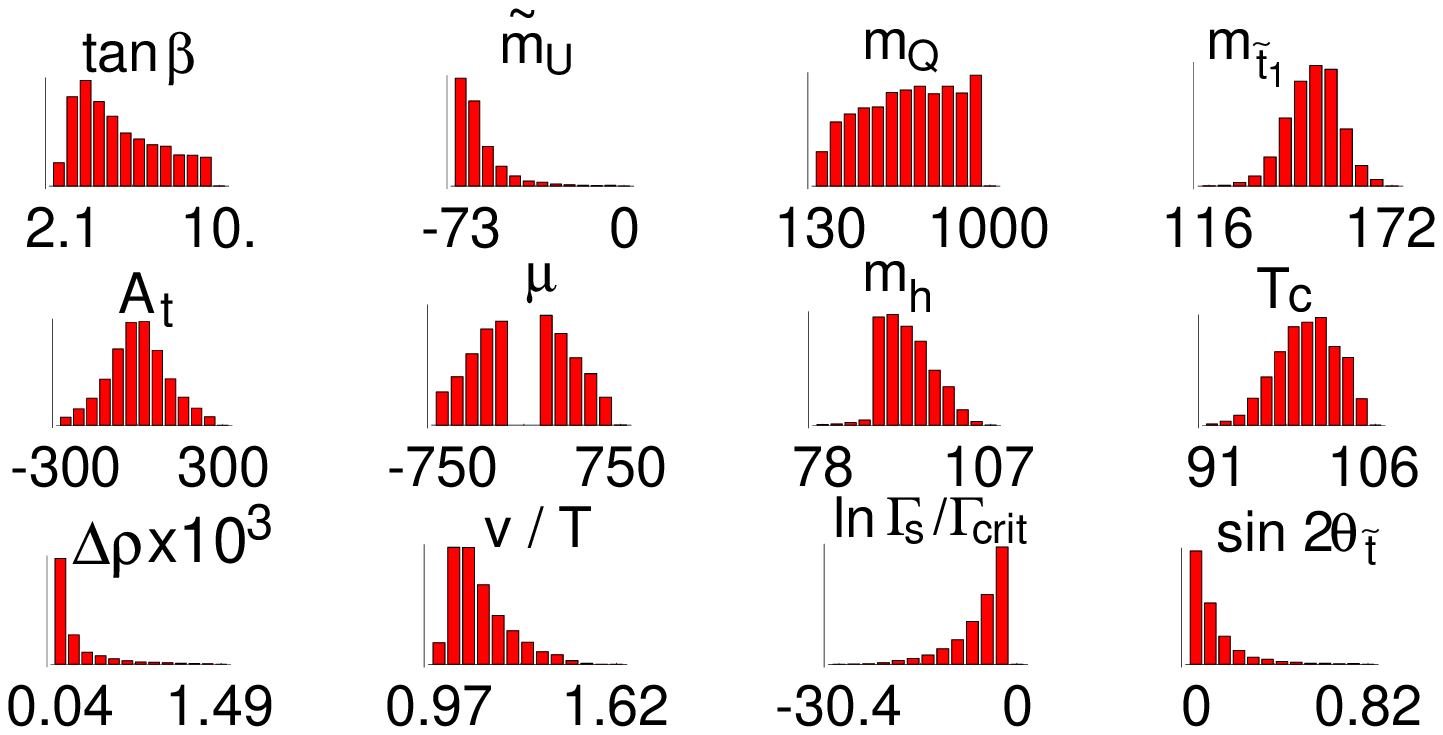}}
\medskip\noindent
{\small Figure 2: Frequencies of baryogenesis-allowed parameters.
Masses are in GeV; $\theta_{\tilde t}$ is
the stop mixing angle.}
\medskip

The strength of the phase transition depends most sensitively on
$\tan\beta$ and $m^2_U$, which in turn determine the masses of the
lightest Higgs boson and the top squarks,
\beqa
	m^2_h &=& \sfrac12\left[m^2_A+m^2_Z-\sqrt{(m^2_A+m^2_Z)^2
	-4m^2_Z m^2_A \cos^2 2\beta}\right]\nonumber\\ 
	&+& O\left((m_t^4/v^2)\ln\left((m_{\tilde t_1}m_{\tilde t_2}
	)/m_t^2\right)\right);\nonumber\\
	{\cal M}_{\tilde t}^2 &\cong& 
	\left(\begin{array}{ll} m^2_Q + m^2_t + O(m^2_Z) &
	m_t(A_t - \mu\cot\beta) \\ m_t(A_t - \mu\cot\beta) &
	m^2_U + m^2_t+ O(m^2_Z)\\ \end{array}
	\right).
\eeqa
Qualitatively, the dependences can be understood as follows.  In the
absence of the light stop, and ignoring the experimental constraint on
the Higgs boson mass, a strong phase transition requires a small value
$m_h$ and hence small $\tan\beta$.  This is because
the quartic terms of the tree-level potential, $g^2(
|H_1|^2-|H_2|^2)^2$, are flat along the direction of $\tan\beta = 1$
($|H_1|=|H_2|$), so the effective quartic coupling $\lambda$ is
minimized for $\tan\beta\sim 1$, which helps the strength of the
phase transition, since $v/T\sim g^3/\lambda$.  The experimental bound on
$m_h$ translates into a lower limit on $\tan\beta$, which excludes the
whole region where baryogenesis is viable.  However, this can be
counteracted if the (mostly right-handed) stop is sufficiently light.
Thus the contours of constant $v/T$ resemble hyperbolas
in the $\tan\beta$-$\tilde m_U$ plane, as
shown in figure 3.  We define $\tilde m_U\equiv m^2_U/|m_U|$ so that
$\tilde m_U$ has the same sign as $m^2_U$.

\centerline{\epsfxsize=3.5in\epsfbox{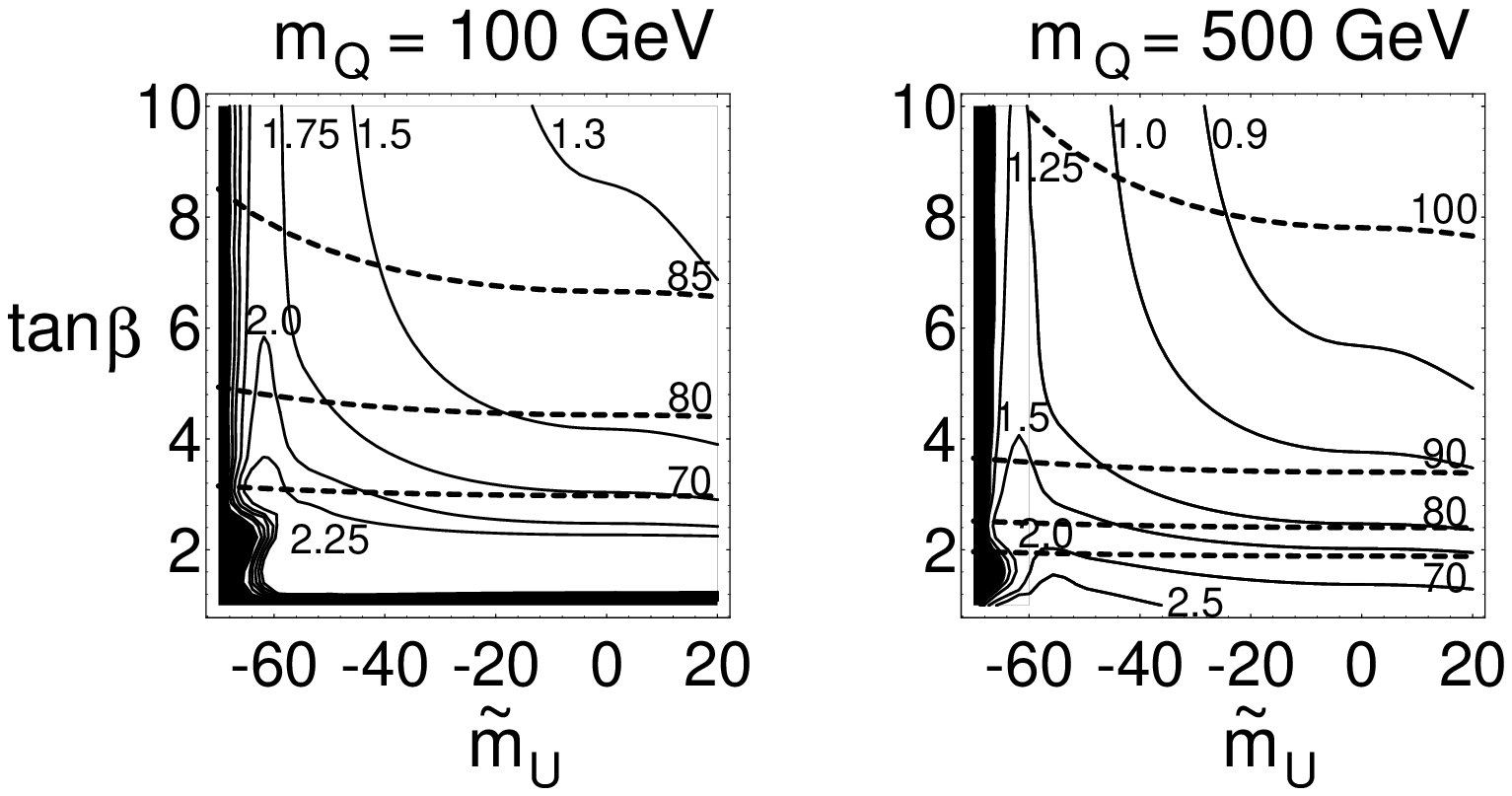}}
\noindent
{\small Figure 3:  Contours of $v/T$ (solid) and Higgs mass (dashed)
in the plane of $\tan\beta$ and $\tilde m_U\equiv m^2_U/|m_U|$ (in GeV), 
for $m_Q = 100$ and 500 GeV, respectively, at zero squark mixing
($\mu=A_t=0$).  The potential has color-breaking minima
in the black regions near $\tilde m_U = -70$ GeV.}
\medskip

The next most important parameter is the soft-breaking mass term for
the left-handed top squark, $m^2_Q$.  It affects the strength of the
phase transition almost exclusively through a radiative correction to
the Higgs self-coupling, proportional to $\ln ( (m^2_Q + m_t^2)/m_t^2
)$.  Thus, at fixed $\tan\beta$, raising $m_Q$ increases $m_h$ and
weakens the phase transition.  (This weakening can be compensated by
sufficiently lowering the right-stop mass.)  However the allowed range
of $\tan\beta$ consistent with the experimental limits on $m_h$
increases with $m_Q$, as shown in figure 4, which has the scatter plots
from the Monte Carlo for $m_h$ and $\tan\beta$ as functions of $\hat
m_Q\equiv m_Q/100$ GeV.  The upper limit on $m_h$ as a function of
$m_Q$ is simply the maximum theoretically allowed value in the MSSM.
The fitted functions for the upper limit on $m_h$ and the corresponding
lower limit on $\tan\beta$ are
\beqa
\label{fits}
	m_h &\lsim & 85.9 + 9.2 \ln(\hat m_Q) \ {\rm GeV}
	\nonumber\\
	\tan\beta &\gsim & 
	(0.03 + 0.076\,\hat m_Q - 0.0031\,\hat m_Q^2)^{-1},
\eeqa
generalizing the findings of ref.~\cite{cqw2}.  As for the smallest
possible values of $m_h$, we see from figure 1 that the scarcity
of points with $m_h < 88$ GeV is due to the small probability of
getting a strong phase transition when $m_A < 100$ GeV.

We have somewhat painstakingly reconsidered the criterion for a strong
enough phase transition.  We first find the nucleation temperature
$T_n$, which is lower than $T_c$, and the latent heat of the phase
transition.  From these we get the temperature to which the universe
reheats, $T_r$, on completion of the phase transition.  We compute the
sphaleron energy using the two loop effective potential at this
temperature, and compare the resulting sphaleron rate $\Gamma_s$ to the
expansion rate of the universe, also accounting for the time dependence
of $\Gamma_s$, eq.~(\ref{sphrate}).  In figure 5a we show the
correlation between the rigorous measure of baryon dilution,
$-\ln\Gamma_s/\Gamma_{\rm crit}$, and $v(T_c)/T_c$.  Here, in contrast
to figure 2, $\Gamma_s/\Gamma_{\rm crit}$ is computed without applying
the lattice correction factor to $v(T_c)/T_c$.  The points below the
line $-\ln\Gamma_s/\Gamma_{\rm crit}=0$ should be discarded, according
to eq.~(\ref{sphrate}), and are only retained if we account for the
fact that the effective potential underestimates $v/T$.  The
correlation between the correct criterion and $v(T_c)/T_c$ is good but
not perfect.  The smallest allowed value of $v(T_c)/T_c$ is 1.05, and
the largest rejected value is about 1.15.

\centerline{\epsfxsize=3.5in\epsfbox{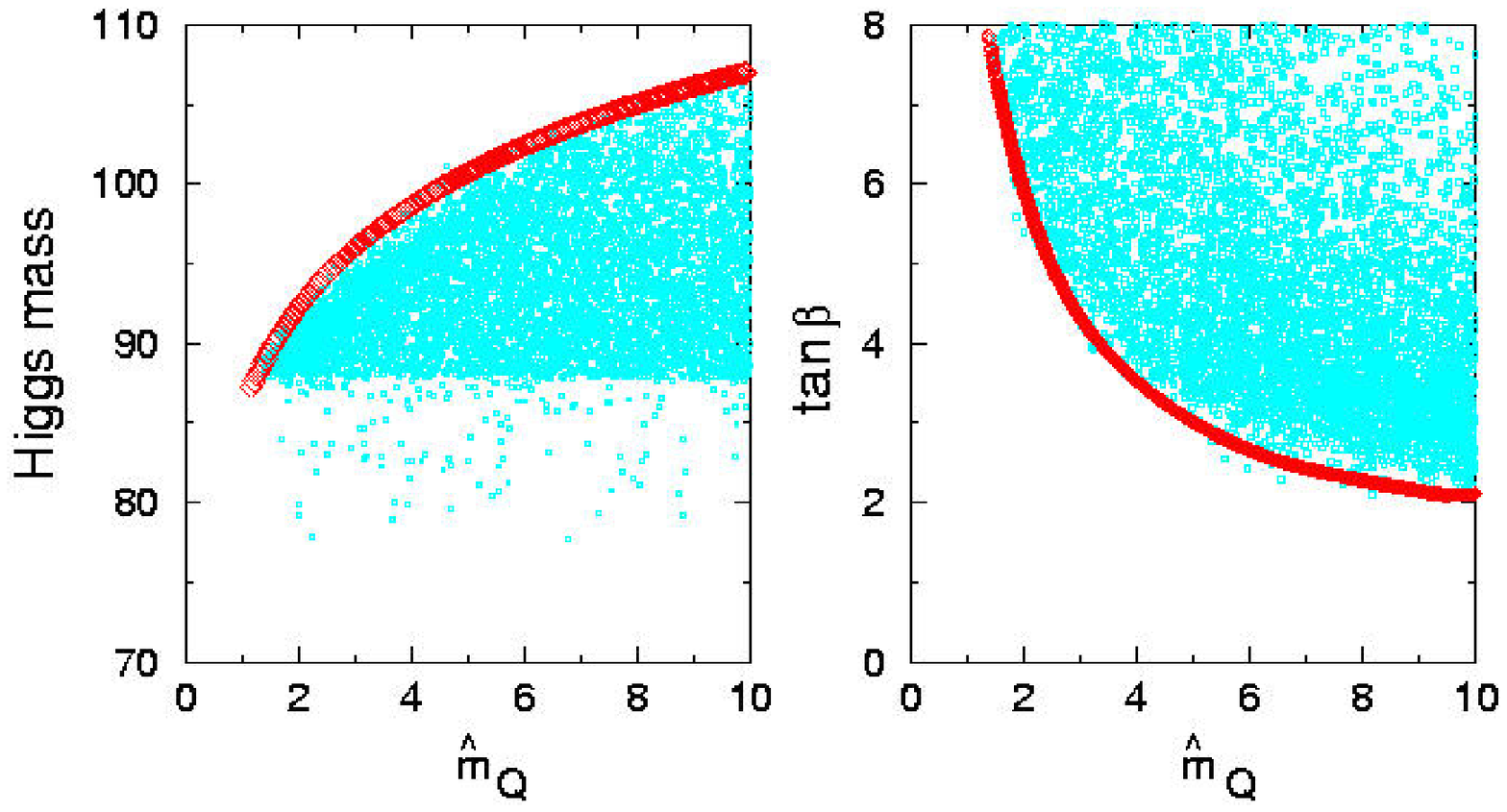}}
\noindent
{\small Figure 4:  Correlation of $m_h$
and $\tan\beta$ with the left-handed top squark mass parameter $\hat
m_Q\equiv m_Q/100$ GeV.  Heavy lines show the approximate limiting
values, eq.~(\ref{fits}). }\medskip

\centerline{\epsfxsize=3.5in\epsfbox{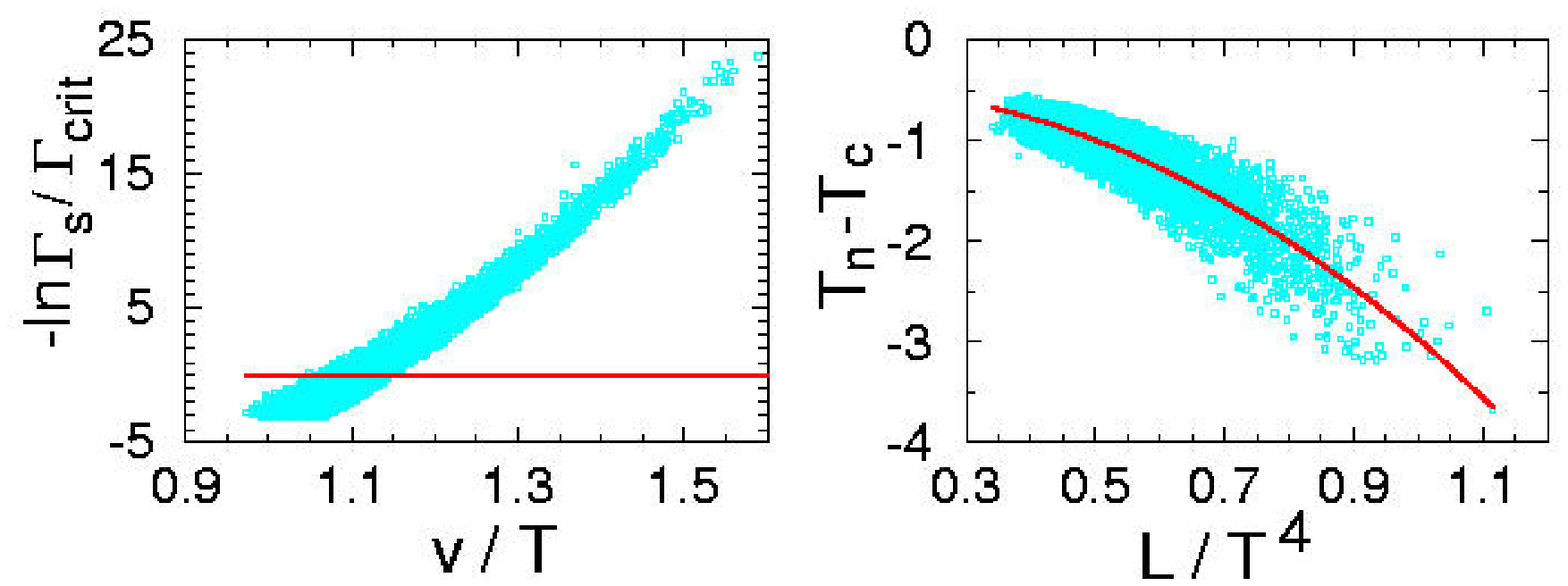}}
\noindent
{\small Figure 5: (a) Correlation of the sphaleron rate with $v/T$;
below the line would be ruled out baryon preservation,
eq.~(\ref{sphrate}); (b) $\Delta T\equiv T_n - T_c$ in GeV versus
$\Lambda \equiv({\rm latent \; heat})/T^4$, fit by the function $\Delta
T = -0.5 + 0.5 \Lambda - 2.9 \Lambda^2$.} \medskip

Another measure of the strength of a first order phase transition is
the latent heat $L$, which is the difference between phases of 
$dV/d\ln T$ at the critical temperature, where $V$ is the effective
potential.  It correlates strongly with the amount of supercooling
before bubble nucleation.  Figure 5b shows the relation between 
$T_n - T_c$ and the latent heat.  The latent
heat also determines the reheat temperature:  $L$ is the heat
available for increasing the plasma energy density, 
\beq
L = \Delta\rho_{r,n}
\simeq g_*\pi^2(T_r^4-T_n^4)/30 \, .
\eeq
Therefore, writing $\Delta\rho_{c,n} = 
g_*\pi^2(T_c^4-T_n^4)/30$, it follows that
\beq
	(T_c-T_r)/(T_c-T_n) \simeq 1 - L/\Delta\rho_{c,n} \, ,
\eeq
and the universe reheats back to $T_c$ if the
ratio $L/\Delta\rho_{c,n}$ ever exceeds unity.  However we find that
$L/\Delta\rho_{c,n}$ has an average value of 0.29 and never falls
outside the range $[0.17, 0.42]$, so reheating to $T_c$ does not
ever occur.

We have also investigated the value of $\tan\beta = v_2/v_1 =
\VEV{H_2}/ \VEV{H_1}$ inside the bubble wall.  This quantity is of
interest because most (but not all \cite{cjk}) electroweak baryogenesis
mechanisms in the MSSM predict the baryon asymmetry is proportional to
an average of $v_2\partial v_1-v_1\partial v_2$ over the the bubble
wall, where $v$ goes from 0 to the broken phase value $v_c$.  We
characterize the variation of $\tan\beta$ with $v$ by minimizing the
effective potential in the angular direction for each fixed value of $v
= (\VEV{H_1}^2+\VEV{H_2}^2)^{1/2}$, and computing $\Delta \beta \equiv
{\rm max}_v[v (\beta(v)-\beta(v_c))]/v_c$.  This definition of
$\Delta\beta$, which differs from ref.~\cite{mqs}'s, corresponds more
closely to the quantity which enters into computations of the baryon
asymmetry.  Like ref.~\cite{mqs}, we find that this quantity never
exceeds $0.02$, and is typically 10 times smaller.  It is also never
large if the $A^0$ boson is heavy.

\centerline{\epsfxsize=3.5in\epsfbox{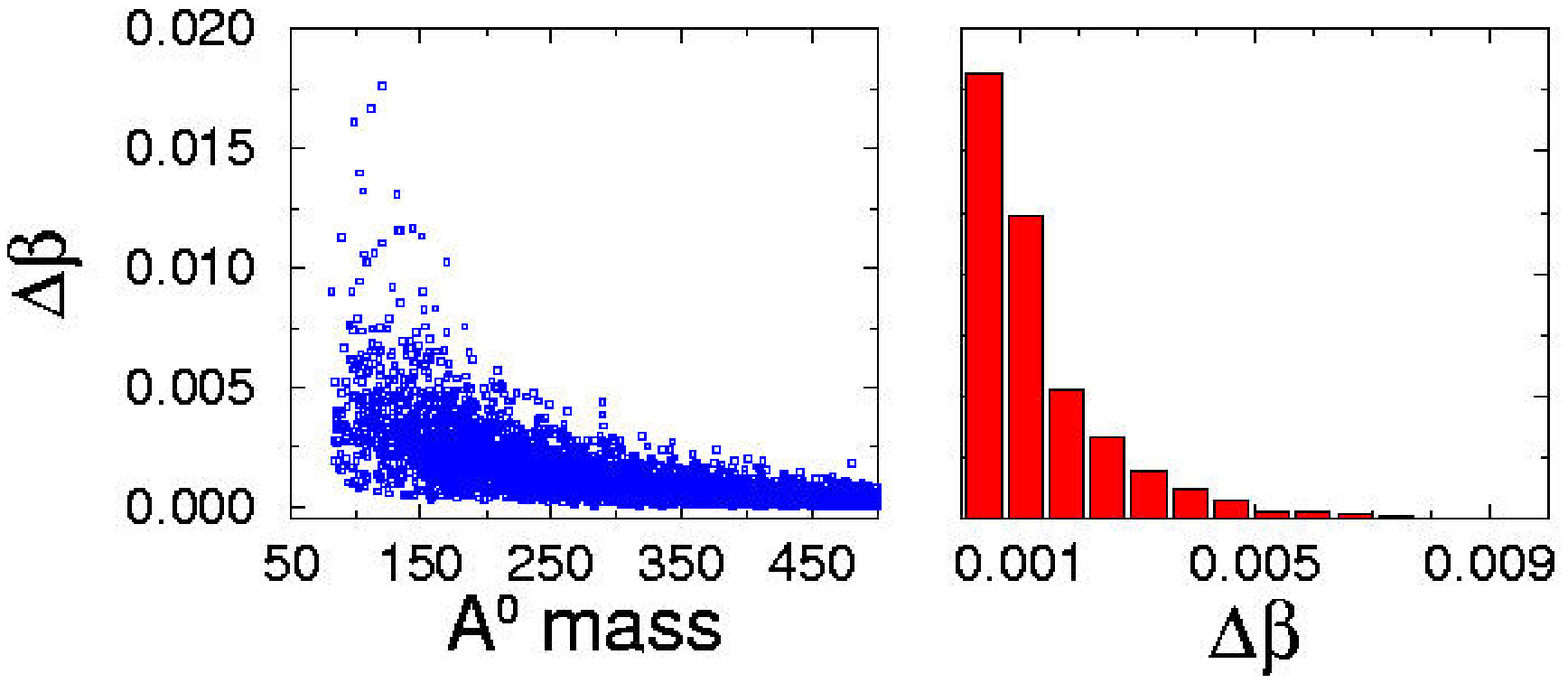}}
\noindent {\small
Figure 6: (a) Maximum deviation in weighted Higgs VEV orientation,
$\Delta\beta\equiv {\rm max}_v[v (\beta(v)-\beta(v_c))]/v_c $, inside
bubble wall, as a function of $m_{A^0}$; (b) distribution of
$\Delta\beta$ values.}\medskip

The most interesting issue confronting electroweak baryogenesis in the
MSSM is whether the LEP 200 run will be able to rule it out.  The
answer seems to be ``almost, but not completely."  The final center of
mass energy $\sqrt{s} = 200$ GeV will exclude $m_h$ up to 107 GeV
\cite{eg}.  However if $m_Q = 2$ TeV, we find parameters with $m_h$ as
high as 116 GeV yet consistent with electroweak baryogenesis.
Such a Higgs boson might be discovered in Run II at Tevatron 
\cite{valls}.

\medskip We wish to thank Jose Ramon Espinosa for patiently helping us
to corroborate the results of references \cite{jre,dce}, Daniel Treille
for clarification of LEP's Higgs boson limits, and Mariano Quiros for
pointing out ref.~\cite{mqs}, which studied bubble nucleation issues
and $\Delta\beta$ prior to this work.



\begin{references}

\bibitem{eqz} J.R.\ Espinosa, M.\ Quiros and 
	F.\ Zwirner,  Phys.\ Lett.\ B307 (1993) 106, hep-ph/9312296
\bibitem{beqz}  A.\ Brignole,  J.R.\ Espinosa, M.\ Quiros and 
	F.\ Zwirner, Phys.\ Lett.\ B324 (1994) 181, hep-ph/9312296 
\bibitem{cqw1} M.\ Carena, M.\ Quiros and C.E.M.\ Wagner, 
	Phys.\ Lett.\ B380 (1996) 81; hep-ph/9603420 
\bibitem{jre} J.R.\ Espinosa, Nucl.\ Phys.\ B475 (1996) 273, hep-ph/9604320 
\bibitem{ck1} J.M.~Cline and K.~Kainulainen, Nucl.~Phys.~B482 (1996) 
    73, hep-ph/9605235 
\bibitem{laine} M.~Laine, Nucl.~Phys.~B481 (1996) 43, hep-ph/9605283
\bibitem{losada} M.~Losada, Phys.~Rev.~D56 (1997) 2893, hep-ph/9605266
\bibitem{fl} G.R.~Farrar and M.~Losada, Phys.~Lett.~B406 (1997) 60,
	hep-ph/9612346 
\bibitem{bjls} D.~B\"odeker, P.~John, M.~Laine and M.G.~Schmidt,
	Nucl.~Phys.~B497 (1997) 387, hep-ph/9612364 
\bibitem{cqw2} M.~Carena, M.~Quiros and C.E.M.~Wagner, 
	preprint FERMILAB-PUB-97-327-T, (1997), hep-ph/9710401 
\bibitem{ck2} J.M.~Cline and K.~Kainulainen, Nucl.~Phys.~B510 (1998) 
	88, hep-ph/9705201 
\bibitem{dce} B.~de Carlos and J.R.~Espinosa, Nucl.~Phys.~B503 (1997),
	hep-ph/9705315 
\bibitem{mqs} J.M.\ Moreno, M.\ Quiros and  M.\ Seco, preprint 
 IEM-FT-168-98 (1998), hep-ph/9801272
\bibitem{lr1} M.~Laine and K.~Rummukainen, preprint CERN-TH-98-121 (1998),
	hep-ph/9804255; preprint CERN-TH-98-122 (1988),
	hep-lat/9804019
\bibitem{hn} P.~Huet and A.E.~Nelson, Phys.~Rev.~D53 (1996) 4578, 
	hep-ph/9506477
\bibitem{cqrvw}  M.\ Carena, M.\ Quiros, A.\ Riotto, I.\ Vilja and 
	C.E.M.\ Wagner, Nucl.\ Phys.\ B503 (1997) 387,  hep-ph/9702409 
\bibitem{aos1} M.\ Aoki, N.\ Oshimo and A.\ Sugamoto, 
	Prog.\ Theor.\ Phys.\ 98 (1997) 1179, hep-ph/9612225 
\bibitem{cjk} J.M.\ Cline, M.\ Joyce and K.\ Kainulainen, Phys.\ Lett.\ 
	B417 (1998) 79, hep-ph/9708393 
\bibitem{kls} A.~Kusenko, P.~Langacker and G.~Segre, Phys.~Rev.~D54 (1996)
	5824, hep-ph/9602414 
\bibitem{gm} G.D.\ Moore, preprint McGILL-97-36 (1998), hep-ph/ 9801204;
	preprint  McGILL-98-7 (1998), hep-ph/9805264
\bibitem{HHG} J.F.~Gunion, H.E.~Haber, G.L.~Kane and S.~Dawson,
{\it The Higgs Hunter's Guide}, Addison-Wesley, Redwood City (1990)
\bibitem{aleph1} ALEPH Collaboration, preprint ALEPH 98-029 (1998)
\bibitem{aleph2} ALEPH Collaboration, preprint ALEPH 98-015 (1998)
\bibitem{abc} G.~Altarelli, R.\ Barbieri and F.\ Caravaglios, 
	Int.\ J.\ Mod.\ Phys.\ A13 (1998) 1031, hep-ph/9712368
\bibitem{opal} OPAL Collaboration, Eur.\ Phys.\ J.\ C2 (1998) 213,
	hep-ex/9708018 
\bibitem{eg} E.~Gross, D.~Lellouch and A.L.~Read, 
 	preprint CERN-EP-98-094 (1998).
\bibitem{valls} J.~Valls, talk at ICHEP `98 meeting, Vancouver, 
	24 July 1988

\end{references}
\end{document}